\documentclass{iaus}
\usepackage{graphicx}

\title[SN II Light Curves from CCCP] 
{The Flavours of SN II Light Curves}

\author[Iair Arcavi]   
{Iair Arcavi$^1$}

\affiliation{$^1$Department of Particle Physics and Astrophysics, The Weizmann Institute of Science, \\ Rehovot 76100, Israel \\ email: {\tt iair.arcavi@weizmann.ac.il}}

\pubyear{2012}
\volume{279}  
\pagerange{xxx--yyy}
\jname{Death of Massive Stars: Supernovae and Gamma-Ray Bursts}
\editors{P. Roming, N. Kawai \& E. Pian, eds.}
\begin{document}

\maketitle

\begin{abstract}
We present $R$-Band light curves of Type~II supernovae (SNe) from the Caltech Core Collapse Program (CCCP). With the exception of interacting (Type~IIn) SNe and rare events with long rise times, we find that most light curve shapes belong to one of three distinct classes: plateau, slowly declining and rapidly declining events. The latter class is composed solely of Type~IIb SNe which present similar light curve shapes to those of SNe Ib, suggesting, perhaps, similar progenitor channels. We do not find any intermediate light curves, implying that these subclasses are unlikely to reflect variance of continuous parameters, but rather might result from physically distinct progenitor systems, strengthening the suggestion of a binary origin for at least some stripped SNe. We find a large plateau luminosity range for SNe~IIP, while the plateau lengths seem rather uniform at approximately $100$ days. We present also host galaxy trends from the Palomar Transient Factory (PTF) core collapse SN sample, which augment some of the photometric results. 
\end{abstract}

\firstsection 
\section{Introduction}

Type~II supernovae (SNe) are widely recognized as the end stages of massive H-rich stars. Together with Type~Ib/c events, they represent the bulk of observed core collapse SNe (see \cite[Filippenko 1997]{filippenko97} for a review of SN classifications). Several sub-types of Type~II SNe have been observed. Those showing a plateau in their light curve are known as Type~IIP events, while those showing a linear decline from peak magnitude are classified as IIL. A third class of events, Type~IIb, characterized by its spectral rather than its photometric properties, develops prominent He features at late times. Finally, Type~IIn SNe display narrow lines in their spectra, indicative of interaction between the SN ejecta and a dense circum stellar medium. 

Red supergiants (RSGs) have been directly identified as the progenitors of Type~IIP SNe (see \cite[Smartt 2009)]{smartt09} for a review. Such stars have thick hydrogen envelopes that are ionized by the explosion shock wave. As the shocked envelope expands and cools, it recombines, releasing radiation at a roughly constant rate, thus producing a plateau in the light curve (e.g. \cite[Popov 1993]{popov93}; \cite[Kasen \& Woosley 2009]{kasenwoosley09}). It follows that SNe IIL might be the explosions of stars with less massive H envelopes that can not support a plateau in their light curve. SN IIb progenitors, then, would contain an even smaller H mass.

However, if SNe IIP-IIL-IIb progenitors represent merely a sequence of decreasing H-envelope mass, one would expect the properties of these SNe to behave as a continuum. Specifically, a gradual transition in light curve shape should be observed when examining a homogeneous sample of events. 

The Caltech Core Collapse Program (CCCP) is a large observational survey which made use of the robotic 60-inch (P60) and Hale 200-inch telescopes at Palomar Observatory to obtain optical $BVRI$ photometry and spectroscopy of 48 nearby core collapse SNe. 

Light curves of Type~Ib/c SNe from CCCP have been presented and analyzed by \cite[Drout et al. (2011)]{drout11}. Type~IIn CCCP events are treated by \cite[Kiewe et al. (2012)]{kiewe12}. Here we present the $R$-Band photometry of 21 non-interacting Type~II SNe with well observed light curves collected through CCCP. 

\section{Photometry}

Our light curves are produced using image subtraction with via the CPM method (\cite[Gal-Yam et al. 2008a]{gy08a}) for PSF matching.
Due to incomplete data for three of the events, we use photometry published in the literature. The light curve of SN2004fx is taken from \cite[Hamuy et al. (2006)]{hamuy2006}, that of SN2005ay from \cite[Gal-Yam et al. (2008b)]{gy08b} and that of SN2005cs from \cite[Pastorello et al. (2009)]{pastorello09}.

\section{Results and Discussion}

\begin{figure}[h!]
\begin{center}
 \includegraphics[width=12cm]{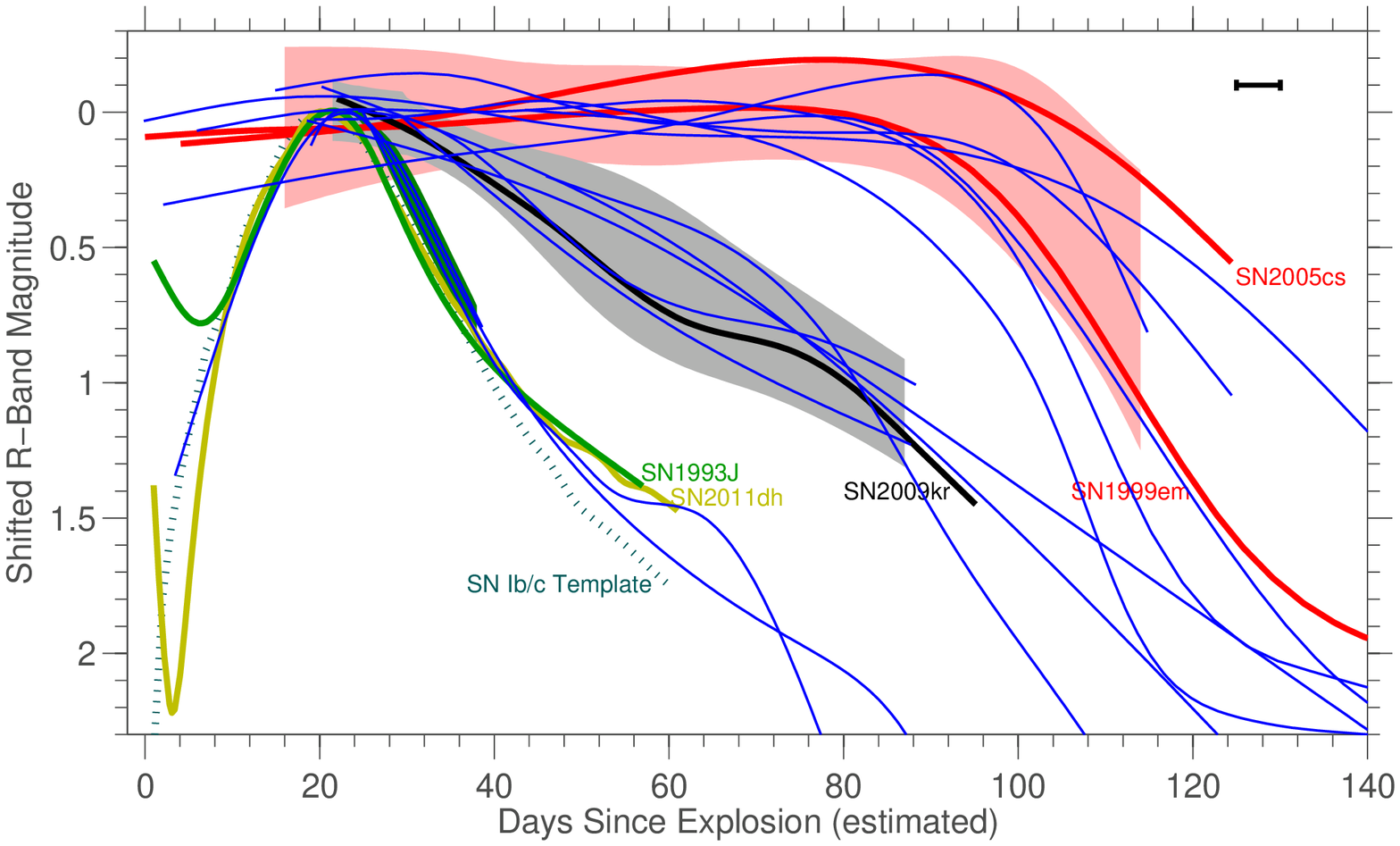} 
 \caption{$R$-band light curves of 15 Type~II SNe from CCCP, normalized in peak magnitude (SN2004fx data taken from \cite[Hamuy et al. 2006]{hamuy2006}; SN2005ay data taken from \cite[Gal-Yam et al. (2008b)]{gy08b}; SN2005cs data taken from \cite[Pastorello et al. 2009]{pastorello09}). Reference SNe are shown for comparison (SN1999em from \cite[Leonard et al. 2002]{leonard02}; SN2009kr from \cite[Fraser et al. 2010]{fraser10}; SN1993J from \cite[Richmond et al. 1994]{richmond94}; SN2011dh from \cite[Arcavi et al. 2011]{arcavi11}). We also overplot the SN Ib/c template derived by \cite[Drout et al. (2011)]{drout11}. The data have been interpolated with spline fits (except for SN2005by, where a polynomial fit provided a better trace to the data). The shaded regions denote the average light curve $\pm2\sigma$ of each subclass. The typical 5-day uncertainty in determining the explosion times is illustrated by the interval in the top right corner.}
   \label{alllcs}
\end{center}
\end{figure}

We plot the $R$-Band light curves of $15$ Type~II events normalized to peak magnitude in Figure \ref{alllcs}. Rather than forming a continuum, we find that the light curves group into three distinct sub-classes: plateau, slowly declining (1-2 Mag/100 days) and initially rapidly declining (5-6 Mag/100 days) events. We note that the three rapidly declining events are all Type~IIb and that they display similar light curve shapes to those of Type~Ib/c SNe (\cite[Drout et al. 2011]{drout11}). 

\begin{figure}[h!]
\begin{center}
\includegraphics[width=12cm]{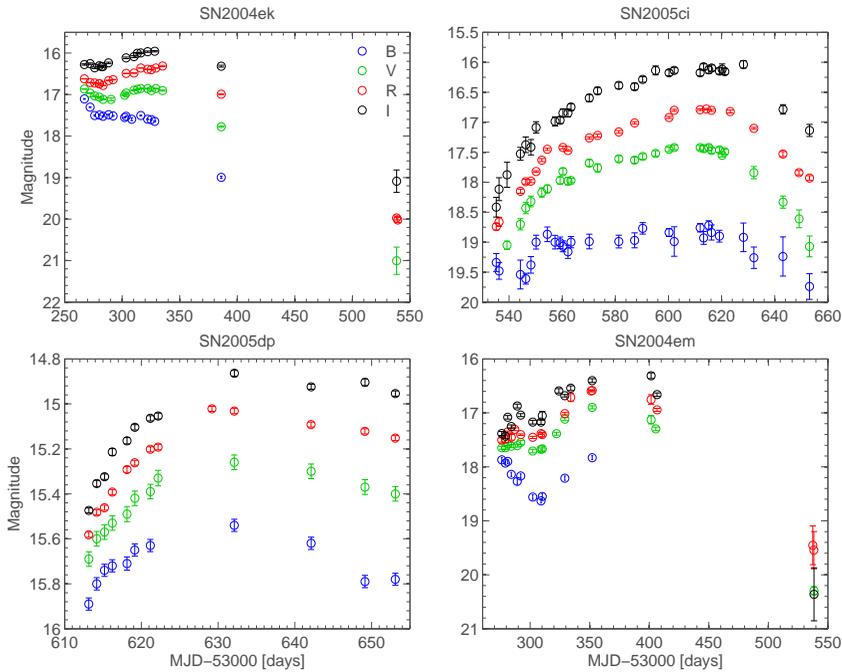}
 \caption{$BVRI$ light curves of the CCCP events not included in Figure \ref{alllcs}. Three events (SN2004ek, SN2005ci, SN2005dp) show long rise times, while one peculiar event (SN2004em) changes behavior from flat to rising around three weeks after explosion.}
   \label{pecs}
\end{center}
\end{figure}

Three events (SN2004ek, SN2005ci and SN2005dp; Figure \ref{pecs}) display prolonged rising periods in their light curves. They do not show signs of interaction in their spectra and may be explosions of compact blue supergiant progenitors (\cite[Kleiser et al. 2011]{kleiser11}; \cite[Pastorello et al. 2012]{pastorello12}). 

\begin{figure}[h!]
\begin{center}
 \includegraphics[width=12cm]{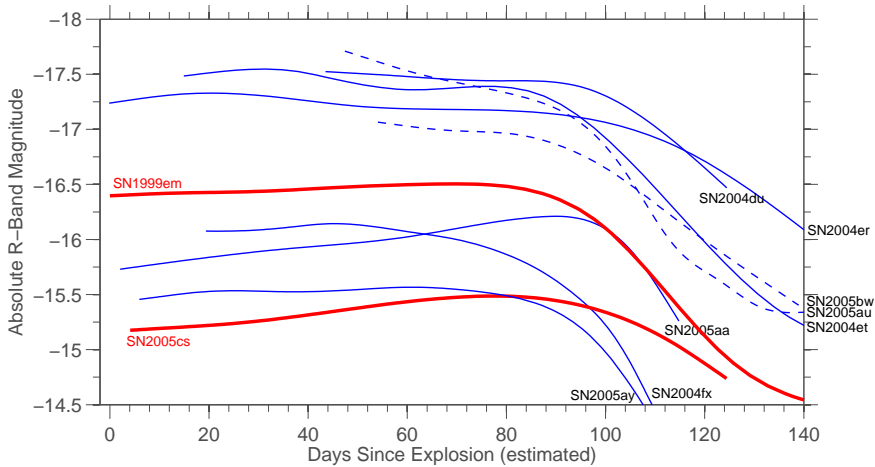} 
 \caption{$R$-band light curves of 9 Type~IIP SNe from CCCP (SN2004fx data taken from \cite[Hamuy et al. 2006]{hamuy2006}; SN2005ay data taken from \cite[Gal-Yam et al. (2008b)]{gy08b}; SN2005cs data taken from \cite[Pastorello et al. 2009]{pastorello09}) with respect to their estimated explosion time (except for SN2005au and SN2005bw, marked by dashed lines, for which the explosion date is not known to good accuracy). SN1999em (\cite[Leonard et al. 2002]{leonard02}) is shown for comparison. Spline fits were applied to the data}
   \label{alllcsiip}
\end{center}
\end{figure}

Finally, one event (SN2004em; Figure \ref{pecs}) displays a very peculiar photometric behavior. For the first few weeks it is similar to a Type~IIP SN, while around day $25$ it suddenly changes behavior to resemble a SN1987A-like event.

\subsection{Declining SNe}

\begin{figure}[h!]
\begin{center}
\includegraphics[width=12cm]{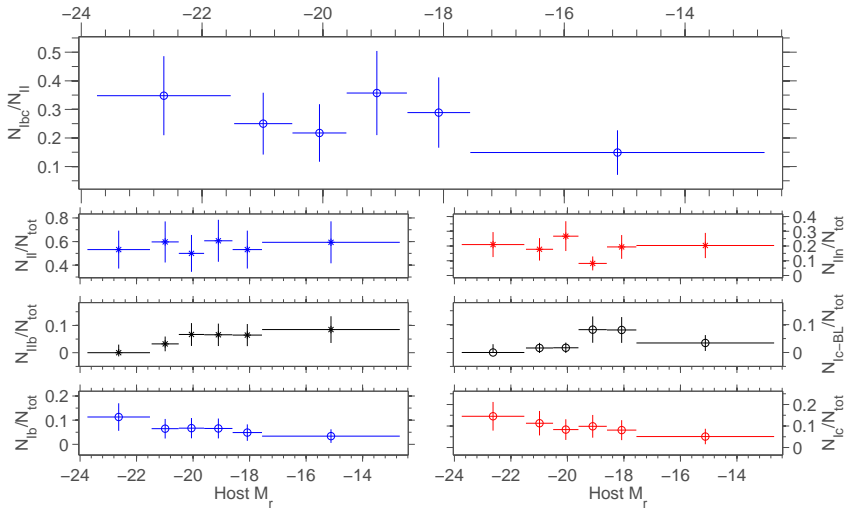}
 \caption{Core collapse SN fractions as a function of host galaxy luminosity, considered as a proxy for metallicity (\cite[Tremonti et al. 2004]{tremonti04}), for 369 events discovered and followed through the Palomar Transient Factory (PTF; \cite[Rau et al. 2009]{rau09}, \cite[Law et al. 2009)]{law09}.}
   \label{metallicity}
\end{center}
\end{figure}

Aside from establishing a different rate of decline for SNe IIb compared to SNe IIL, Figure \ref{alllcs} suggests that the IIP, IIL and IIb subtypes do not span a continuum of physical parameters, such as H envelope mass. Rather, additional factors should be considered. Specifically, Type~IIb events might arise from binary systems (as suggested also by recent progenitor studies for SN1993J, \cite[Maund et al. 2004]{maund04}; SN2008ax, \cite[Crockett et al. 2008]{crockett08}; SN2011dh, \cite[Arcavi et al. 2011]{arcavi11}, \cite[Van-Dyk et al. 2011]{vd11}). The similarity of the Type~IIb light curves to those of Type~Ib events (in addition to the known spectral similarities at late times) suggests that these two types of events might come from similar progenitor systems. Metallicity might be an important factor driving some of the systems to explode as Ib SNe and others as IIb's (Fig. \ref{metallicity}; Arcavi et al., in prep.).

\subsection{Plateau SNe}

The $R$-Band light curves of the Type~IIP SNe, on an absolute magnitude scale, can be seen in Figure \ref{alllcsiip}. We find a wide range of plateau luminosities, but do not have enough statistics to test whether they form a continuous distribution or if there are two distinct underlying types (bright and faint), as previously suggested (\cite[Pastorello et al. 2004]{pastorello04}). The plateau lengths, however, seem rather uniform at $\sim100$ days (with the sole exception of SN2004fx, displaying a shorter plateau)\footnote{note that SN2005au and SN2005bw are plotted only to show their plateau luminosity, their plateau lengths are unknown due the lack of sufficient constraints on their explosion time}. 

The scarcity of observed short plateaus and the possibly related sharp distinction between IIP and IIL light curves is evident. Such a pronounced absence of intermediate events might suggest that Type~IIL SNe are powered by a different mechanism than that associated with SNe~IIP (e.g., \cite[Kasen \& Bildsten 2010]{kasebildsten10}).

\begin{discussion}

\discuss{Chornock}{Can you comment on the peak luminosities of the objects you have classified as SNe IIL / slow decliners?}

\discuss{Arcavi}{We do not have the peaks resolved for all these events, but they seem to be confined to the $-16$ to $-17$ magnitude range.}

\discuss{Chornock}{What about SN 1979C?}

\discuss{Arcavi}{SN 1979C, known as the canonical Type~IIL, does not fit into any of the three subtypes we identify in Fig. \ref{alllcs}. SN 1979C was exceptionally bright and it is still debated whether it was an interacting event. We therefore choose not to inlcude it in the analysis.}

\discuss{Smith}{You showed different behavior of SNe IIb vs. Ib with host galaxy luminosity (and presumably metallicity). Have you looked for a similar trend with the location within the host galaxy?}

\discuss{Arcavi}{The location within the host galaxy can serve as a proxy for progenitor mass, and is therefore and excellent complementary paramater to check. We hope to be able to conduct this analysis in the future, especially for the very faint host galaxies in the PTF sample (most of which require the resolution of \it{HST}).} 

\discuss{Katz}{Is there a contious spectrum between SNe Ib and IIb?}

\discuss{Arcavi}{Yes, in fact at late times, the two events have identical spectra (which is why multi-epoch spectroscopy is required to correctly classify them).}

\end{discussion}

\end{document}